\documentstyle[psfig]{mn}

\newcommand{\kms}	{km~s$^{-1}$}
\newcommand{\h}   	{$h^{-1}\,$~kpc}
\newcommand{\etal} 	{{et~al.}}
\newcommand{\push}[1]	{\multicolumn{1}{c}{#1}}
\newcommand{\hst}	{{\it{HST}}}

\title[Redshifts of galaxies close to QSOs]
{Redshifts of galaxies 
close to bright QSO lines of sight}

\author[D.V. Bowen et al.]
{David V. Bowen,$^{1,2}$
\thanks{Visiting astronomer, Kitt Peak National Observatory, National
Optical Astronomy Observatories, operated by the Association of Universities
for Research in Astronomy, Inc., under cooperative agreement with the National
Science Foundation}
Samantha J. Osmer,$^2$ J. Chris Blades,$^2$ and
David Tytler$^3$\\
$^1$ Royal Observatory, Edinburgh, Blackford Hill, Edinburgh EH9 3HJ\\
$^2$ Space Telescope Science Institute, 3700 San Martin Drive, Baltimore,
MD 21218, U.S.A\\
$^3$ Center for Astrophysics and Space Sciences, University of 
California, San Diego, CASS 0111, La Jolla, CA 92093-0111, U.S.A.
}

\begin{document}

\maketitle

\begin{abstract}

To expand the known number of low-redshift galaxies which lie close to bright
($V\:<\:17.2$) QSO lines of sight, we have identified 24 galaxies within 11
arcmins of nine QSOs which have been observed with the {\it Hubble Space
Telescope} (\hst ).  Galaxies are found between redshifts of $0.0114-0.145$
and lie between $39-749$~\h\ from QSO sightlines.  Knowing the redshifts of
these galaxies has already proved important in understanding results from
\hst\ programmes designed to search for UV absorption lines from low-redshift
galaxies, and will enable future observations to probe the halos of these
galaxies in detail.

\end{abstract}


\section{INTRODUCTION}

Finding low-redshift galaxies close to high-redshift QSOs on the plane of the
sky is important for probing the interstellar medium (ISM) of the disks and
halos of present epoch galaxies, since the light of the background QSO may be
absorbed by gas along the line of sight.  Both optical and ultraviolet (UV)
absorption lines which arise in nearby galaxies have been studied previously
[see Bowen, Blades \& Pettini (1995; hereafter BBP) and refs therein]; in
particular, the detection of UV lines is important not only for understanding
the physics of galaxy ISMs other than that of the Milky Way, but because these
are the same lines observed out to redshifts of $\sim 4$ in QSO spectra
obtained from ground-based observatories. The intention is to find low
redshift analogs to the high redshift systems, which will enable us to
understand the origin and evolution of absorbing gas from the earliest epochs.

Previous compilations of QSO-galaxy
pairs (Monk et~al.~1986; Burbidge~et~al.~1990) 
have collated well known cases of QSO-galaxy pairs and added additional data
where available. Often though, not listed in these catalogs are
other (fainter) galaxies closer to the QSO sightline which can also be probed
once their redshifts have been determined. More seriously,
searching for UV absorption lines from nearby galaxies requires the use of
the {\it Hubble Space Telescope} (\hst ).  Unfortunately, 
the most severe constraint in studying galaxies in this way is the small
number of QSOs
which are bright enough
(particularly in the far UV where the
satellite's spectrographs are less sensitive) {\it and} close enough to a
galaxy to be of interest. Hence finding galaxies near to UV bright QSOs is
extremely important.

In this paper we present the redshifts of 26 objects which lie within 11
arcmins of nine QSO lines of sight. Three QSOs were originally observed with
\hst\ as part of BBP's Mg~II survey; the field around these QSOs clearly
showed galaxies other than those which were being probed, but which had no
known redshifts.  The lines of sight to two QSOs were also discussed by Bowen,
Blades, \& Pettini (1996) in a \hst\ Archival programme designed to search for
Ly$\alpha$ absorption from low-redshift galaxies.  The remaining QSOs were
observed as part of the {\it HST} Faint Object Spectrograph Quasar Absorption
System Snapshot Survey (`AbSnap') programme (Bowen~et~al.~1994;
Tytler~et~al. 1996).  A subsample of the AbSnap target fields
showed numerous faint galaxies close to the
QSO lines of sight in the 
digitised images 
collated by Bowen~et~al.~(1994).
These galaxies are interesting because although the
exposure times for the AbSnap QSOs were short, the QSOs are some of the
brightest available.

\section{OBSERVATIONS}

Galaxies were identified from 16.7 arcmins square images extracted from
the STScI Digitized Sky Survey (DSS),
although the galaxies nearest a QSO sightline were selected
preferentially for spectroscopic follow up. Magnitudes of the galaxies
could not be calculated due to the saturation of the digitised images at
magnitudes of $V\:\leq\:15$ and a lack of photometry of any fainter
objects from which to calculate a reliable photometric zero point.
The $V$ magnitude limit for these images is $\sim\:19.5$ (Golombeck 1993), so
all galaxies are brighter than this limit.

Reproduction of parts of the DSS images of interest are reproduced in
Figures~1$-$5.  Table~1 lists the QSO fields studied (col.~1), and the QSO's
redshift and $V$ magnitude taken from V\'{e}ron-Cetty \& V\'{e}ron (1993)
(cols.~3 and 2). The positions of galaxies for which spectra were obtained are
given in column~5. Galaxies labelled in Figures~$1-5$ are identified in
column~4, while separations between QSO and galaxy on the plane of the sky in
arcmins, $\rho$, are given in column~6.

\begin{table*}
\centering
\caption{New Redshifts of galaxies close to \hst\ target QSOs}
\begin{tabular}{cccllrrcr}

    &     &                & Galaxy & \multicolumn{2}{c}{Galaxy RA \& DEC}
& \push{$\rho$} &      & \push{$\rho '$}       \\

QSO & $V$  & $z_{\rm{QSO}}$ & \push{ID}     & \multicolumn{2}{c}{(1950.0)} 
& \push{($'$)}  & $z$  & \push{($h^{-1}$ kpc)} \\

(1) & (2) & (3)            & (4)    & \multicolumn{2}{c}{(5)}      
& \push{(6)}    & (7)  & \push{(8)}            \\\hline 

Q0026+1259 & 15.4 & 0.142 & G4 & 00 26 39.68 & 13 04 21.2 & 4.88 & 0.0394 &
158.3 \\

           &      &       & G5 & 00 27 04.40 & 13 02 49.5 & 7.23 & 0.1453 &
749.1 \\

Q0100+0205 & 16.4 & 0.394 & G12 & 01 00 15.20 & 02 05 17.0 & 8.92 & 0.0444 &
323.6 \\

Q0122$-$0021 & 16.7 & 1.070 & G1 & 01 22 55.43 & $-$00 19 02.4 & 2.51 & 0.0213 &
45.2 \\

Q0219+4248 & 15.2 & 0.444 & G1$^{\rm{a}}$ & 02 19 21.60 & 42 50 16.8 & 2.37 & 0.0198 &
39.7 \\

           &      &       & G2 & 02 19 18.32 & 42 47 09.2 & 2.53 & 0.0667 &
133.6 \\

           &      &       & G3 & 02 19 40.76 & 42 50 10.2 & 2.59 & $\ldots$      & $\ldots$ \\

           &      &       & G4 & 02 19 41.18 & 42 45 37.6 & 3.52 & 0.0674 &
187.7 \\

           &      &       & G5$^{\rm{b}}$ & 02 19 48.91 & 42 47 04.4 & 3.74 & 0.0219 &
69.1 \\

           &      &       & G6 & 02 19 48.37 & 42 45 15.7 & 4.66 & 0.0675 &
248.8  \\

           &      &       & G7 & 02 19 54.13 & 42 41 08.1 & 8.58 & 0.0230 &
166.3 \\

           &      &       & G8 & 02 19 39.85 & 42 53 10.9 & 5.03 & 0.0184 &
78.5 \\

Q1544+4855 & 16.5 & 0.400 & G1 & 15 43 56.89 & 48 56 00.0 & 0.76 & 0.0746 &
44.4 \\

           &      &       & G2 & 15 43 56.24 & 48 54 11.7 & 1.40 & 0.0753 &
82.5 \\

           &      &       & G5 & 15 43 49.72 & 48 57 06.6 & 2.38 & 0.0382 &
75.0 \\

           &      &       & G7 & 15 43 48.96 & 48 56 33.1 & 2.13 & 0.0 & $\ldots$ \\

           &      &       & G8 & 15 43 57.61 & 48 54 32.3 & 0.99 & 0.0 & $\ldots$ \\

Q1631+3930 & 16.0 & 1.023 & G4 & 16 30 43.70 & 39 37 21.3 & 9.57 & 0.0291 &
232.6 \\

Q1701+6102 & 17.0 & 0.164 & G1 & 17 01 31.22 & 61 01 59.8 & 1.04 & 0.0659 &
54.3 \\

           &      &       & G2 & 17 01 32.08 & 61 03 22.7 & 0.47 & 0.0451:  & 20.5  \\
           &      &       & G3 & 17 01 32.12 & 61 03 33.8 & 0.63 & 0.0451   & 27.5 \\

           &      &       & G4$^{\rm{c}}$ & 17 02 27.05 & 61 06 44.1 & 7.43 & 0.0114 &
72.6 \\

Q2215$-$0347 & 17.2 & 0.241 & G1 & 22 15 12.29 & $-$03 47 27.5 & 0.23 & $\ldots$ & $\ldots$ \\

           &      &       & G4 & 22 14 57.85 & $-$03 48 20.1 & 3.64 & 0.0615 &
178.6 \\

           &      &       & G5 & 22 14 59.64 & $-$03 50 03.9 & 3.94 & $\ldots$ & $\ldots$ \\

           &      &       & G6 & 22 14 44.01 & $-$03 45 47.8 & 7.29 & 0.0560 &
328.2 \\

           &      &       & G7 & 22 14 40.83 & $-$03 54 43.3 & 10.53 & 0.0568 &
480.3 \\

           &      &       & G10 & 22 14 39.13 & $-$03 54 19.3 & 10.59 & 0.0574:
& 487.7 \\

Q2349$-$0125 & 15.3 & 0.174 & G4 & 23 49 08.27 & $-$01 17 44.7 & 8.90 & 0.0385 &
282.4 \\
\hline
\multicolumn{9}{l}{$^{\rm{a}}$ UGC 1832; $^{\rm{b}}$ UGC 1837; $^{\rm{c}}$ NGC
6292}
\end{tabular}
\end{table*}

Spectroscopic observations of candidate galaxies were made using GOLDCAM and a
FORD chip on the KPNO 2.1~m telescope between $2-6$ September 1993.  A 2
arcsec slit aligned north-south was used with the 300 lines mm$^{-1}$ grating
blazed at 6750~\AA\ to obtain galaxy spectra with a resolution of 7.2~\AA .
The total wavelength range covered was $\simeq\:3800-7500$~\AA .  
Where possible, galaxy redshifts were measured by cross-correlating the 
galaxy spectra with
template spectra of two radial velocity standards, HD~171391
and HD~182572 
using the {\tt fxcor} routine in IRAF. For some galaxies, only emission lines
of H$\alpha$, [NII], and [SII] were visible, and for these objects redshifts
were measured by fitting gaussians to the emission lines.  The resulting
redshifts are given in column~7 of Table~1---a blank entry means that a
spectrum was obtained of the object but no redshift could be measured due to
the low signal-to-noise of the spectrum.  The error in the redshifts is
$\Delta z \approx \pm 0.0003$.  The corresponding distance of the galaxy from
the QSO sightline, $\rho '$, is given in column~8 in units of \h , where
$h\:=\:H_{0}/100$, $H_{0}$ is the Hubble constant, and $q_{0}\:=\: 0$.

\section{RESULTS}

Below, we outline the results for each of the fields studied, and draw
attention to what is already known about the absorbing properties of some of
the galaxies. For each sightline studied, we have also searched the NASA
Extragalactic Database (NED) for any other known low-redshift galaxies within
0.5~$h^{-1}$~Mpc from the QSO line of sight.  
No galaxies are found in this way unless otherwise stated in \S3.1$-$3.9 below.

\subsection{Q0026+1259}

Galaxy G4 in Figure~\ref{pics1}a (G0026+1304) was originally identified by
Monk~\etal\ (1986) to be at a redshift of $z\:=\:0.0058$, placing it only
28~\h\ from the QSO line of sight.  However, H$\alpha$, [N~II] and [S~II] are
clearly identified at a redshift of $z\:=\:0.0394$.  This means that the
separation between galaxy and QSO sightline is 158~\h .  No Ly$\alpha$
absorption is found from G4 to an equivalent width limit of 0.65~\AA\ (Bowen,
Blades \& Pettini 1996).  G5 (G0027+1302) is an irregularly shaped galaxy at
the same redshift as the QSO, $z\:=\:0.1453$, 749~\h\ from the QSO line of
sight.

\begin{figure*}
\vspace*{22cm}
\caption{\label{pics1}
Portions of the `Quick-$V$' POSS plates extracted from the Digitised
Sky Survey (DSS) images available at STScI, showing the galaxies close to
(a) Q0026+1259 (top) and (b) Q0100+0205 (bottom)
for which spectra were obtained.
North East is to the top left for all images. Scales are different for
each image, and are shown in the bottom right corner.}
\end{figure*}

\subsection{Q0100+0205}

We have measured the redshift of galaxy G12 in
Figure~\ref{pics1}b to be $z\:=\:0.0444$. The galaxy is 8.92 arcmins from the
QSO line of sight which corresponds to $\rho\:=\:324$~\h . This
galaxy is not the closest known identified object however: galaxy
G11 is UGC0656 (at RA$\:=\:01:01:10.48$
DEC$\:=\:02:00:54$, a more accurate position than that given in the NED). Its
position corresponds to a separation of 9.0 arcmins from the QSO line of
sight, or $\rho'\:=\:140.6$~\h\ at a redshift of $z\:=\:0.01841$. Even closer,
however, is the dwarf irregular galaxy, IC~1613. The sightline to the QSO
passes $\approx\:28.5$ arcmins from the galaxy; assuming a distance of
$\sim\:0.7$~Mpc to IC~1613 (Huterer et al. 1996, and refs therein) this
separation corresponds to 5.8~kpc. The sightline to Q0100+0205 will therefore
provide a valuable probe of the outer regions of the interstellar medium of a
dwarf galaxy for future \hst\ observations.

\subsection{Q0122$-$0021}

This QSO line of sight passes 2 degrees, or 
1.8~$h^{-1}$~Mpc, from the center of Abell~194 (Chapman~et al.~ 1988).  
There are three galaxies at the cluster velocity which lie within 500~\h\ 
of the QSO sightline, UGC~0998, CGCG~385$-$115, and CGCG~385$-$141, at 
separations of 13.9, 20.5 and 22.4 arcmins, respectively, which correspond 
to 201, 326, and 353~\h\ at the galaxies' redshifts.  The galaxy whose 
redshift we have measured, labelled G1 in Figure~\ref{pics2}a, 
is at a velocity of 
$cz = 6386$~\kms .  The velocity dispersion of the Abell~194 is $\simeq\: 
450$~\kms , which means that if G1 is a member of the cluster, its velocity 
is $2\sigma$ away from the cluster's systemic velocity.

Whether or not G1 is a member of Abell~194, it lies 2.51 arcmins from the
QSO line of sight, which corresponds to only 44.8~\h . There are three other
galaxies which lie between Abell~194 and the Milky Way and which are within
0.5~$h^{-1}$~Mpc of the QSO line of sight: UGC~1011, UM~323, and UGC~0931,
which, with velocities of 1923, 1799, and 1948~\kms , and separations of 26.4,
38.1 and 49.3 arcmins, pass 146, 198 and 277~\h\ from the QSO sightline.

\subsection{Q0219+4248}

The galaxies close to the line of sight of Q0219+4248 previously identified by
Arp (1968) are members of Abell~347. Galaxy G1 labelled in Figure~\ref{pics2}b
is Galaxy D measured by Arp, and our redshift of $z\:=\:0.0198$ measured from
absorption lines agrees well with his value of $z\:=\:0.0200$.  G5 is Arp's
Galaxy C, and we measure $z\:=\:0.0219$, identical to Arp's original value.
More details on the environment of the galaxies, and
the search for Mg~II along the line of sight, can be found in BBP.

In addition to confirming the redshifts of G1 and G5, we have also obtained
redshifts of 6 other galaxies. G7 and G8 are probably also members of
Abell~347 with $z\:=\:0.0230$ and $z\:=\:0.0184$ respectively. However, we
have found three galaxies (G2, G4 and G6) with redshifts of
$z\:\simeq\:0.067$, all within 300~\h\ of the sightline to Q0219+4248. It
therefore seems likely that the QSO line of sight intercepts a second cluster
behind Abell~347, and there is already evidence for the existence of
Ly$\alpha$ absorption at this cluster velocity (Bowen, Blades \& Pettini
1996).

\begin{figure*}
\vspace*{22cm}
\caption{\label{pics2}Same as Figure 1, for (a) Q0122$-$0021 (top) and (b)
 Q0219+4248 (bottom).}
\end{figure*}

\subsection{Q1544+4855}

G1 (G1543+4856) and G2 (G1543+4854) were originally identified by Monk~\etal\
(1986) although no redshifts were available. We have confirmed the redshifts
measured by Bowen~\etal\ (1991), and in addition, we have shown that two of
the objects close on the plane of the sky are stars (Figure~\ref{pics3}a).  The
existence of Mg~II absorption at the velocity of these two galaxies has been
discussed in detail by BBP, who postulate that the absorption arises from gas
distributed away from the galaxies as a result of galaxy-galaxy
interactions. An $R$-band CCD image of the field can also be found in BBP.  G5
is actually at a lower redshift than the G1/G2 pair, at $z\:=\:0.0382$, so
although it is nearly twice as far on the sky than G2, the separation to the
line of sight is only 75~\h .

\subsection{Q1631+3930}

We find only one low-redshift galaxy clearly visible near Q1631+3930. G4 in
Figure~\ref{pics3}b is 9.57 arcmins from the QSO sightline, and has a redshift
of 0.0291. The separation therefore corresponds to 232.6~\h . A search in the
NED finds few galaxies with known redshifts close to the line of sight; only
IC~4610 with $cz\:=\:9260$~\kms\ lies within 0.5~$h^{-1}$~Mpc, 12.7 arcmins,
or 339~\h\ away.

\begin{figure*}
\vspace*{22cm}
\caption{\label{pics3}Same as Figure 1, for (a) Q1544+4855 (top) and (b) 
Q1631+3930 (bottom).}
\end{figure*}

\begin{figure*}
\vspace*{22cm}
\caption{\label{pics4}Same as Figure 1, for (a) Q1701+6103 (top) and (b) 
Q2215$-$0347 (bottom).}
\end{figure*}

\begin{figure*}
\vspace*{22cm}
\caption{\label{pics5}Same as Figure 1, for Q2349$-$0125.}
\end{figure*}

\subsection{Q1701+6102}

The field around Q1701+6102 is dominated by NGC~6292, labelled G4 in
Figure~\ref{pics4}a which lies 7.43 arcmins away from the QSO line of
sight. We measure its velocity to be $cz\:=\:3418$~\kms , which places it
72.6~\h\ from the QSO sightline. This redshift measurement is at odds with
that measured optically by Bottinelli~et.~al~(1993; listed in the NED) who
found $cz\:=\:3805\pm70$~\kms. However, our redshift, measured from strong
H$\alpha$, [N~II], and [S~II] lines, agrees well with an earlier 21~cm H~I
measurement by Richter and Huchtmeier (1991), who found $cz\:=\:3407$~\kms .
Closer to the QSO on the plane of the sky is G1, which is only 1.04 arcmins
away, an irregularly shaped galaxy whose redshift we find to be 0.0659, which
places it only 54~\h\ from the QSO line of sight.

The closest galaxies we identify to the QSO line of sight are labelled G2 and
G3 in Figure~\ref{pics4}a. The spectra are of low signal to noise, but a
redshift for G3 of $z = 0.0541$ is clear from the identification of H$\alpha$,
[NII], and [SII] emission lines. G2 is even fainter, and few lines are
discernable in a spectrum with even poorer signal-to-noise than G3. However,
there is an emission line at the same wavelength as the H$\alpha$ emission
line in the spectrum of G3. We therefore tentatively conclude that G2 and G3
are a pair of galaxies at the same redshift, $z = 0.0541$.

This field is also of interest because of the proximity of Q1701+6102 to
another QSO, Q1704+6048 ($\equiv\:$3C~351), 23 arcmins away. Bowen, Blades and
Pettini (1996) searched for Ly$\alpha$ absorption from five galaxies lying
within $124-270$~\h\ of the QSO line of sight, but found none to $3\sigma$
equivalent width limits of 0.32~\AA . Four of these galaxies also lie within
500~\h\ of Q1701+6102, and although their separations from the sightline are
larger, $359-408$~\h , this sightline may prove useful in determining whether
Ly$\alpha$ absorbing gas is distributed only sparsely around groups of
galaxies.

\subsection{Q2215$-$0347}

The field around Q2215$-$0347 shows many bright galaxies, two of which,
labelled G4 and G7 in Figure~\ref{pics4}b, show signs of tidal disturbances,
at least in the DDS image. If the galaxies are disturbed, it is unlikely that
they are interacting with each other, since G4 has a redshift of 0.0615, while
G7 has a redshift of 0.0568, $\sim\:1300$~\kms\ apart. Figure~\ref{pics4}b
shows that there are many other galaxies in the field, for which we were able
to measure the redshifts of two, G6 and G10. Both have redshifts close to G7,
but that of G10 is based on a {\it a priori} knowledge of the redshifts of the
other galaxies in the field and the measurement of Na~I alone, close to that
predicted from the redshifts of the other galaxies. Its redshift is therefore
highly uncertain and is labelled as such in Table~1.

The NED lists only one other galaxy within 0.5~$h^{-1}$~Mpc of Q2215$-$0347,
that of MCG-01-56-005, 15.7 arcmins on the plane of the sky, or 221~\h\ at the
galaxy's velocity of 4969~\kms .

\subsection{Q2349$-$0125}

We find only one bright galaxy close to the sightline of Q2349$-$0125,
labelled G4 in Figure~\ref{pics5}. This galaxy lies 8.90 arcmins away and has a
redshift of 0.0385, which means that it passes 282~\h\ from the QSO line of
sight.

\section{SUMMARY}

We have presented new redshifts for 26 objects which lie close to nine QSO
lines of sight, 24 of which are galaxies.  The galaxies span a redshift range
of 0.0114$-$0.145 and lie $39-749$~\h\ from the QSO line of sight, although
most have $z < 0.1$ and are found within several hundred \h\ of the sightline,
the goal originally set for this programme. The QSOs themselves have $V$-band
magnitudes of $15.4-17.2$, and are therefore excellent sources with which to
probe the intervening galaxies with \hst ; several have already been
succesfully utilized to search for Mg~II or Ly$\alpha$ absorption, and it is
hoped that further observations of all the QSOs will yield valuable insights
into the nature of galaxy disks and halos.

\section*{ACKNOWLEDGMENTS}
It is a pleasure to thank Di Harmer at KPNO for her help with GOLDCAM on the
2.1 m at KPNO.  This research has made use of the NASA/IPAC Extragalactic
Database (NED) which is operated by the Jet Propulsion Laboratory, Caltech,
under contract with the National Aeronautics and Space Administration.  IRAF
is distributed by the National Optical Astronomy Observatories, which is
operated by the Association of Universities for Research in Astronomy, Inc.,
under cooperative agreement with the National Science Foundation.  The work
described in this paper was funded from the following grants: GO-2553.01-87A,
GO-3755.01-91A, and GO-3525.01-91A.


\begin{thebibliography}{}

\bibitem{b1} Arp, H. A. 1968, PASP, 80, 129

\bibitem{b2} Bottinelli, L., et al. 1993, A\&AS, 102, 57

\bibitem{b3} Bowen, D. V., \& Blades, J. C., \& Pettini, M. 1996, ApJ, in press.

\bibitem{b4} Bowen, D. V., \& Blades, J. C., \& Pettini, M. 1995, ApJ, 448,
634 (BBP)

\bibitem{b5} Bowen, D. V., Pettini, M., Penston, M. V., \& Blades, J. C.
1991, MNRAS, 249, 145

\bibitem{b6} Bowen, D. V., Osmer, S. J., Blades, J. C., Tytler, D., Cottrell,
L., Fan, X-M., \& Lanzetta, K. M. 1994, AJ, 107, 461

\bibitem{b7} Burbidge, G., Hewitt, A., Narlikar, J. V., \& Dasgupta, P. 1990,
ApJS, 74, 675

\bibitem{b8} Chapman, G. N. F., Geller, M. J., and Huchra, J. P. 1988, AJ, 95, 999

\bibitem{b9} Golombeck, D., 1993, GASP Cookbook, STScI.

\bibitem{c1} Huterer, D., Sasselov, D. D., and Schechter, P. L. 1995, AJ, 110, 2705

\bibitem{c2} Monk, A. S., Penston, M. V., Pettini, M., \& Blades, J. C. 1986,
MNRAS, 222, 787

\bibitem{c3} Richter, O-G., and Huchtmeier, W. K. 1991, A\&AS, 87, 425


\bibitem{c4} Tytler, D., Zuo, L., Fan, X-M., Bowen, D. V., Blades, J. C. \& Cottrell, L. 
1996, in preparation

\bibitem{c5} V\'{e}ron-Cetty, M.-P. \&  V\'{e}ron, P. 1993, ESO Scientific Report No. 13

\end{thebibliography}
\end{document}